\documentclass[preprint,amsmath,amssymb]{revtex4}
\pdfoutput=1

\usepackage{color} 
\definecolor{darkblue}{rgb}{0,0,0.5}
\definecolor{lila}{rgb}{0.3,0,0.3}
\definecolor{turq}{rgb}{0,0.1,0.4}

\usepackage{url} 
\usepackage[pdftex,
  colorlinks=true,
  linkcolor=darkblue, 
  filecolor=red,
  citecolor=turq, 
  urlcolor=lila, 
  pdftitle={Efficient coupling of photons to a single molecule and the observation of its resonance fluorescence},
  pdfauthor={Gert Wrigge, Ilja Gerhardt, Jaesuk Hwang, Gert Zumofen, Vahid Sandoghdar},
  pdfsubject={Resonance Fluorescence},
  pdfkeywords={Single Molecules, Resonance Fluorescence, Extinction, Absorption, Mollow Triplet, Solid Immersion Lens, SIL},
  pdfpagelabels=true,
  breaklinks=true,
  plainpages=false,
  bookmarks, bookmarksnumbered=true]{hyperref}

\usepackage{graphicx} 
\usepackage{bm}

\begin{document}
\newcommand{\ethaffil}{Laboratory of Physical Chemistry and optETH, ETH Zurich, CH-8093 Zurich, Switzerland}
\newcommand{\real}{\operatorname{Re}}

\title{Efficient coupling of photons to a single molecule and the observation of its resonance fluorescence}

\author{G. Wrigge}
\author{I. Gerhardt}
\author{J. Hwang}
\author{G. Zumofen}
\author{V. Sandoghdar}\email{vahid.sandoghdar@ethz.ch}

\affiliation{Laboratory of Physical Chemistry and optETH, ETH Zurich, CH-8093 Zurich, Switzerland}

\begin{abstract}
Single dye molecules at cryogenic temperatures display many
spectroscopic phenomena known from free atoms and are thus promising
candidates for fundamental quantum optical studies. However, the
existing techniques for the detection of single molecules have
either sacrificed the information on the coherence of the excited
state or have been inefficient. Here we show that these problems can
be addressed by focusing the excitation light near to the absorption
cross section of a molecule. Our detection scheme allows us to
explore resonance fluorescence over 9 orders of magnitude of
excitation intensity and to separate its coherent and incoherent
parts. In the strong excitation regime, we demonstrate the first
observation of the Mollow triplet from a single solid-state emitter.
Under weak excitation we report the detection of a single molecule
with an incident power as faint as 150 attoWatt, paving the way for
studying nonlinear effects with only a few photons.
\end{abstract}

\maketitle

\begin{figure}[b!]
\centering
\includegraphics[width=8 cm]{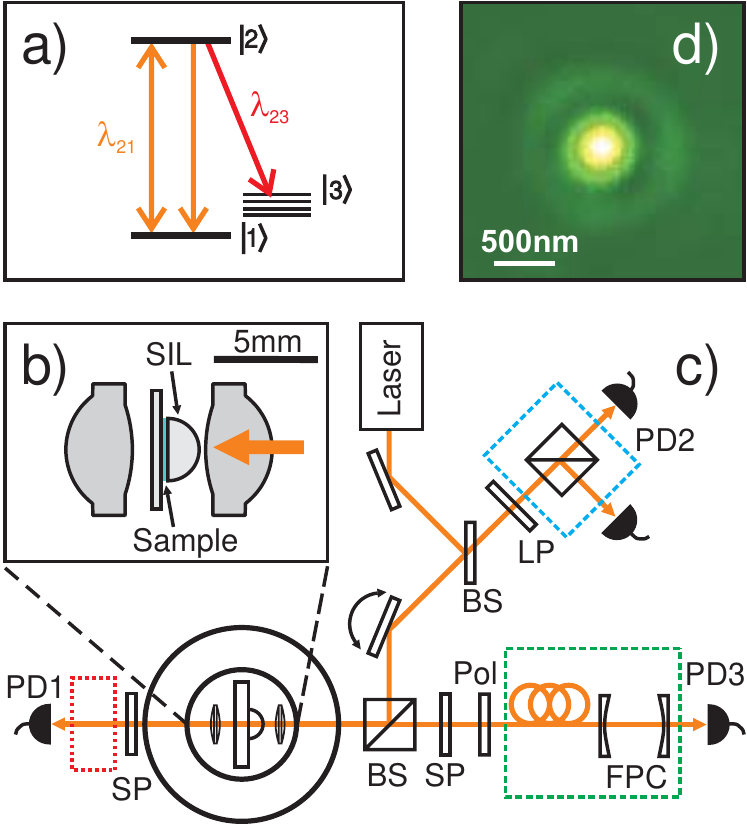}
\caption{\label{setup}a) The energy level scheme of a molecule with ground state
$|1\rangle$ and excited state $|2\rangle$. Manifold $|3\rangle$
shows the vibrational levels of the electronic ground state. b) The
arrangement of the lenses in the illumination and collection paths
in the cryostat. c) The schematics of the optical setup. The dotted
boxes are put in place for specific measurements discussed in the
text. Abbreviations not explained in the text are BS: beam splitter,
SP: short-pass filter, LP: long-pass filter, and Pol: polarizer. d)
A laser scan image of a single molecule, showing a full width at
half-maximum focus spot of about 370~nm.}
\end{figure}

Single dye molecules embedded in solids have been shown to behave as
quantum mechanical two-level systems at cryogenic temperatures,
exhibiting phenomena such as antibunching~\cite{Basche:92b}, AC
Stark shift~\cite{Tamarat:95} and various nonlinear
effects~\cite{Tamarat:99} known from atoms. The pioneering detection
of single molecules was achieved via absorption spectroscopy
assisted by a double lock-in method~\cite{Moerner:89}. In the weak
excitation regime, the effect of a plane-wave illumination on a
molecule with a transition between levels $|1\rangle$ and
$|2\rangle$ at wavelength $\lambda_{21}$ can be formulated as
$I_{\rm d}=(1-\sigma/F) I_{\rm e}$. Here $I_{\rm d}$ is the detected
intensity of the excitation beam after having encountered the
emitter, $\sigma=3\lambda_{21}^2/2\pi$ is the absorption cross
section of the transition, and $F$ and $I_e$ are the area and
intensity of the excitation beam, respectively. The experimental
realization of this simple scenario is challenging because the
effect of a single molecule has to be registered on top of the noise
associated with a considerable laser power and scattering
background. To improve the signal-to-noise ratio (SNR) in single
molecule detection, fluorescence excitation spectroscopy was
introduced~\cite{Orrit:90}. In this technique the broad Stokes
shifted fluorescence at $\lambda_{23}$ is detected on a very low
background after filtering out light at $\lambda_{21}$ (see
Fig.~\ref{setup}a). Unfortunately however, the impressive SNR of
this method comes at the cost of losing the information on the
coherences of the transition at $\lambda_{21}$.

Two recent reports have shown that it is also possible to detect
single molecules coherently~\cite{Plakhotnik:01,Gerhardt:07a} via
interference between the excitation laser beam and the light
scattered by the molecule. In the first scheme, the coherent
scattering and a fraction of the reflected excitation beam were
detected through a $10~\rm{\mu m}$ pinhole~\cite{Plakhotnik:01},
resulting in a ratio of the order of $10^{-6}$ between the
excitation and detected intensities. In the second experiment,
cryogenic scanning near-field optical microscopy (SNOM) was employed
to confine the excitation light to a subwavelength aperture with a
typical transmission of $10^{-5}$~\cite{Gerhardt:07a}. The great
shortcoming of these methods is that more than 100,000 photons are
required to detect one excitation. In view of quantum optical
operations between photons and emitters, it would be highly
desirable to approach a point where as many of the incident photons
as possible interact with a single emitter. Such a regime would open
the door to a wealth of nonlinear interactions between a single
emitter and a single or few photons, which have been so far only
achieved using sophisticated high finesse
microcavities~\cite{Turchette:95,Hijlkema:07,Press:07}. In this
work, we show that it is possible to reach an efficient coupling
between a single emitter and a \emph{freely} propagating laser beam
in a single-pass encounter if one reduces $F$ close to the
diffraction limit~\cite{vanEnk:01}.

\section{Efficient far-field detection of a single molecule in
transmission}

A key technological hurdle in the study of single molecules in the
condensed phase is to achieve a tight focus in the cryostat. Here we
show that this problem can be addressed by exploiting the
solid-immersion lens (SIL) technology~\cite{Mansfield:90}. As shown
in Fig.~\ref{setup}b, a molded glass aspheric lens with $NA=0.68$
and a working distance of 1.76~mm focused the laser beam onto the
sample, and a second identical lens was used to re-collimate it in
the forward direction. A cubic zirconia half-ball lens with a
diameter of 3~mm and a refractive index of about 2.18 was used as a
SIL. Piezoelectrically driven slip-stick sliders were used to adjust
the lateral and axial positions of the aspheric lenses with respect
to the SIL. Fig.~\ref{setup}d shows a confocal fluorescence
excitation scan of a single molecule. We achieved a focal spot with
a full width at half-maximum of about 370~nm at $T=1.4~K$. Although
this still does not meet the diffraction limit, it constitutes a
great improvement over previous cryogenic far-field resolutions and
allowed us to perform several new experiments, as described in the
following.

Figure~\ref{setup}c sketches the experimental arrangement. A tunable
dye laser with a linewidth of 1~MHz was used to excite the
molecules. The laser beam could be scanned across the sample with
the aid of galvanometric mirrors. A Berek variable waveplate was
used to pre-compensate for a slight elliptical polarization that was
accumulated along the optical path and to ensure linear polarization
at the position of the molecule. The samples consisted of a
Shpol'skii matrix doped with dibenzanthanthrene (DBATT)
molecules~\cite{Boiron:96}. To prepare a sample, a small amount of
DBATT was dissolved in \emph{n}-tetradecane, applied to the flat
side of the solid immersion lens and then squeezed and held in place
by a glass cover slip. The 0-0 vibronic transition of DBATT in
\emph{n}-tetradecane is at a wavelength of $\lambda_{21}\sim590$~nm.
We routinely measured near to natural lifetime limited linewidths of
17-20MHz consistent with our independent measurements of 9.7~ns for
the excited state fluorescence lifetime. Another favorable feature
of DBATT is a very short triplet state lifetime~\cite{Boiron:96}.

Figure~\ref{setup}c shows three detection channels that were used
for various measurements discussed in this article. The avalanche
photodiode PD1 recorded the intensity of the transmitted light close
to the excitation wavelength $\lambda_{21}$ by using a short-pass
filter. The second detector (PD2) was situated after a long-pass
filter and registered the Stokes shifted fluorescence on transitions
$2\rightarrow3$. A third detector (PD3) was used in combination with
other optical elements to record the resonance fluorescence
intensity in reflection. In each experimental run, single molecules
were first identified and characterised on PD2 using conventional
fluorescence excitation spectroscopy. After optimization of the
lateral focus position (see Fig.~\ref{setup}d), the polarization of
the excitation light was aligned with the dipole moment of the
molecule. The molecules that were selected for the experiments
presented here showed no significant axial component in the
orientation of their dipole moments and fluoresced with nearly
perfect linear polarization.

\begin{figure}[b!]
\centering
\includegraphics[width=8 cm]{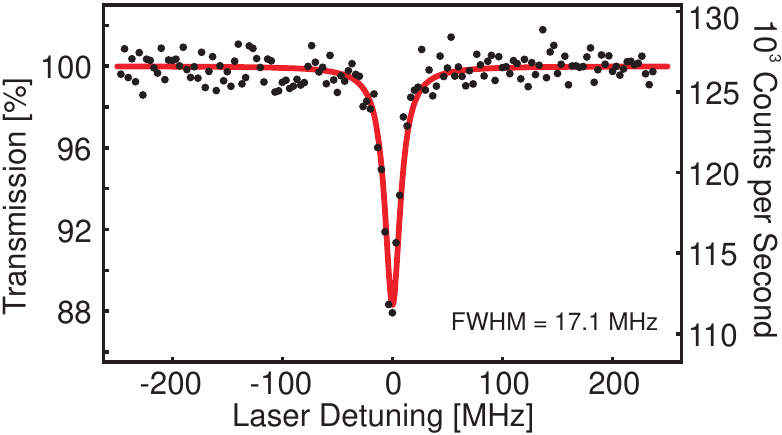}
\caption{\label{dip}An example of a raw transmission spectrum, revealing a
$11.5\%$ dip. The integration time per pixel was 160~ms and the
noise amounts to $0.7\%$.}
\end{figure}

Figure~\ref{dip} shows an example of the raw transmission spectra
recorded on PD1 with the above-mentioned approach. A dip of $11.5\%$
is observed directly in transmission when a single molecule is
resonant with the laser beam. The noise level amounts to 0.7\%,
corresponding to the shot-noise limit of the incident intensity over
an integration time of 160~ms per pixel. We emphasize that no noise
suppression technique such as lock-in detection was employed in this
work.

Although one might be inclined to attribute the observed dip to
``absorption", the excitation spectrum $I_{\rm d}(\nu)$ registered
on the detector has to take into account the interference between
the excitation field $\mathbf{E}_{\rm e}$ and the field
$\mathbf{E}_{\rm m}$ scattered by the
molecule~\cite{Plakhotnik:01,Gerhardt:07a} according to
\begin{equation}
I_{\rm d}=\left\langle \widehat{\mathbf{E}}_{\rm e}^{-}\cdot\widehat{\mathbf{E}}_{\rm e}^{+}\right%
\rangle +\left\langle \widehat{\mathbf{E}}_{\rm
m}^{-}\cdot\widehat{\mathbf{E}}_{\rm m}^{+}\right\rangle +2
\real{\left\{ \left\langle \widehat{\mathbf{E}}_{\rm e}^{-}\cdot\widehat{\mathbf{E}}%
_{\rm m}^{+}\right\rangle \right\}}.
\end{equation}
In our recent work using SNOM~\cite{Gerhardt:07a,Gerhardt:07b} we
expressed the detector signal in detail, accounting for the modal
evolution of the incident and scattered fields and for the
spectroscopic properties of the molecule. Here we simplify the full
expression from Ref.~\cite{Gerhardt:07a} in the form
\begin{eqnarray}
I_{\rm d}&=&I_{\rm e}\left[1+ A\mathcal{L}\left( \nu \right) -
B\mathcal{L}\left( \nu \right) \left( \Delta \cos \psi +
\frac{\gamma }{2}\sin \psi\right)\right]. \label{f-g}\end{eqnarray}
where
\begin{equation}
\mathcal{L}(\nu)=\frac{1}{\Delta ^{2}+\gamma ^{2}/4+\Omega
^{2}(\gamma /2\gamma _{0})}
\end{equation} is a Lorentzian line
profile, and $\Delta$ denotes the laser frequency detuning. The
quantities $\gamma$ and $\gamma_0$ stand for the full width at
half-maximum values of the homogeneous and natural linewidths of the
zero-phonon component of the 0-0 transition, respectively. The
parameter $\Omega=\mathbf{E}_{\rm e} \cdot \langle 1 | \mathbf{D} |
2 \rangle / h$ is the Rabi frequency, where $\mathbf{D}$ signifies
the dipole transition operator. $A$ and $B$ are positive
coefficients whereas $\psi$ signifies the phase difference between
the excitation and the coherently scattered light components.

Equation~(\ref{f-g}) describes the spectrum that one registers on a
detector as one scans the laser frequency. The first term is simply
the part of the incident intensity that arrives at the detector. The
second term proportional to $A$ represents the molecular emission
and is always positive. In total, it consists of the narrow-band
resonance fluorescence at $\lambda_{21}$~\cite{Loudon,cohen-book},
the phonon wing of this transition caused by the vibrational
coupling of the molecule with the matrix, and the Stokes-shifted
fluorescence emitted at $\lambda_{23}$. In the signals detected on
PD1 and PD3, however, the latter is filtered out. The third term in
this equation proportional to $B$ denotes the interference between
$\mathbf{E}_{\rm e}$ and the coherent part of $\mathbf{E}_{\rm m}$.
In Refs.~\cite{Gerhardt:07a,Gerhardt:07b}, we showed various
examples where $\psi$ could be varied, resulting in dispersive
transmission spectra. In the weak excitation limit, the transmission
signal on detector PD1 is dominated by the interference component,
also known as the \emph{extinction} term.

We emphasize that the observed interference effect depends on the
mode overlap between the molecular emission and the incident beam
and can be influenced by the choice of the detection solid angle. In
our case, total internal reflection at the asymmetric geometry of
the SIL-sample-glass-air layers can lead to a fine structure in the
reflected beam~\cite{Novotny:01} and take away energy from the
transmitted beam. This arrangement also results in a strongly
asymmetric molecular emission pattern which favors emission into the
SIL side by about ten times~\cite{Koyama:99}. Furthermore, for a
given geometry the spectroscopic features of the emitter play an
imporant role. The emission associated with the narrow-band
zero-phonon line of the 0-0 vibronic transition is lowered by the
Debye-Waller and Franck-Condon factors $\alpha_{DW}$ and
$\alpha_{FC}$, which amount to 0.25 and 0.3 in our
system~\cite{Gerhardt:07a}, respectively. Thus, as compared to a
perfect two-level system in the same geometric configuration, our
molecule has had $1/(\alpha_{DW}\alpha_{FC}) \sim 12$ times weaker
coherent interaction with the laser beam.

\section{Coherent resonance fluorescence}

\begin{figure}[b!]
\centering
\includegraphics[width=8 cm]{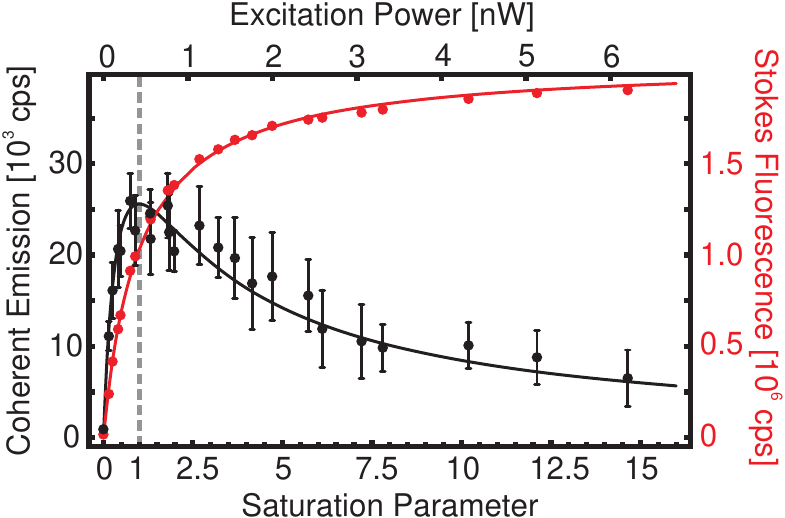}
\caption{Stokes shifted signal (red symbols) and the coherent part
of the resonance fluorescence (black symbols) as a function of the
laser excitation power, also expressed in terms of the saturation
parameter $S$. The solid curves are the results of the theoretical
calculations according to text-book formulae discussed in the
text.\label{coh-incoh}}
\end{figure}

In addition to providing a strong detection signature, direct
extinction spectroscopy offers the great advantage that it gives
access to the narrow-band emission on the zero-phonon line. The
radiation of a two-level system at the wavelength of the excited
optical transition is known as resonance fluorescence. In case of
weak resonant excitation and in the absence of dephasing, this
emission is purely elastic and coherent~\cite{Loudon,cohen-book}. As
the incident intensity is raised, this component first grows
linearly, but it reaches a maximum at saturation parameter
$S=\frac{\gamma\Omega^2 /2\gamma_0}{\Delta^2+\gamma^2/4}=1$. Under
strong illumination, the population of the excited state undergoes
Rabi oscillations at frequency $\Omega$ and the emitted field is
contaminated by an inelastic incoherent component~\cite{cohen-book}.
The coherent part of the resonance fluorescence is predicted to
decrease slowly according to the text-book formula $S/(1+S)^2$,
shown by the solid black curve in Fig.~\ref{coh-incoh}.

The experimental assignment of the coherent and incoherent parts of
resonance fluorescence is a nontrivial task because for a given
excitation intensity, the detected spectrum contains both. The
Lorentzian peak of the $A$-term in Eq.~(\ref{f-g}), receives
contributions from both components. However, only the coherent
emission can interfere with the incident light to produce the
$B$-term, which is in turn composed of a Lorentzian part and a
dispersive signal. Thus, one cannot decipher the variation of the
coherent fluorescence with the excitation intensity in an
unambiguous manner by simply monitoring the dip in Fig.~\ref{dip}.
Below we show that it is possible to do this by influencing the
individual components in a controlled manner.

\begin{figure}[b!]
\centering
\includegraphics[width=8 cm]{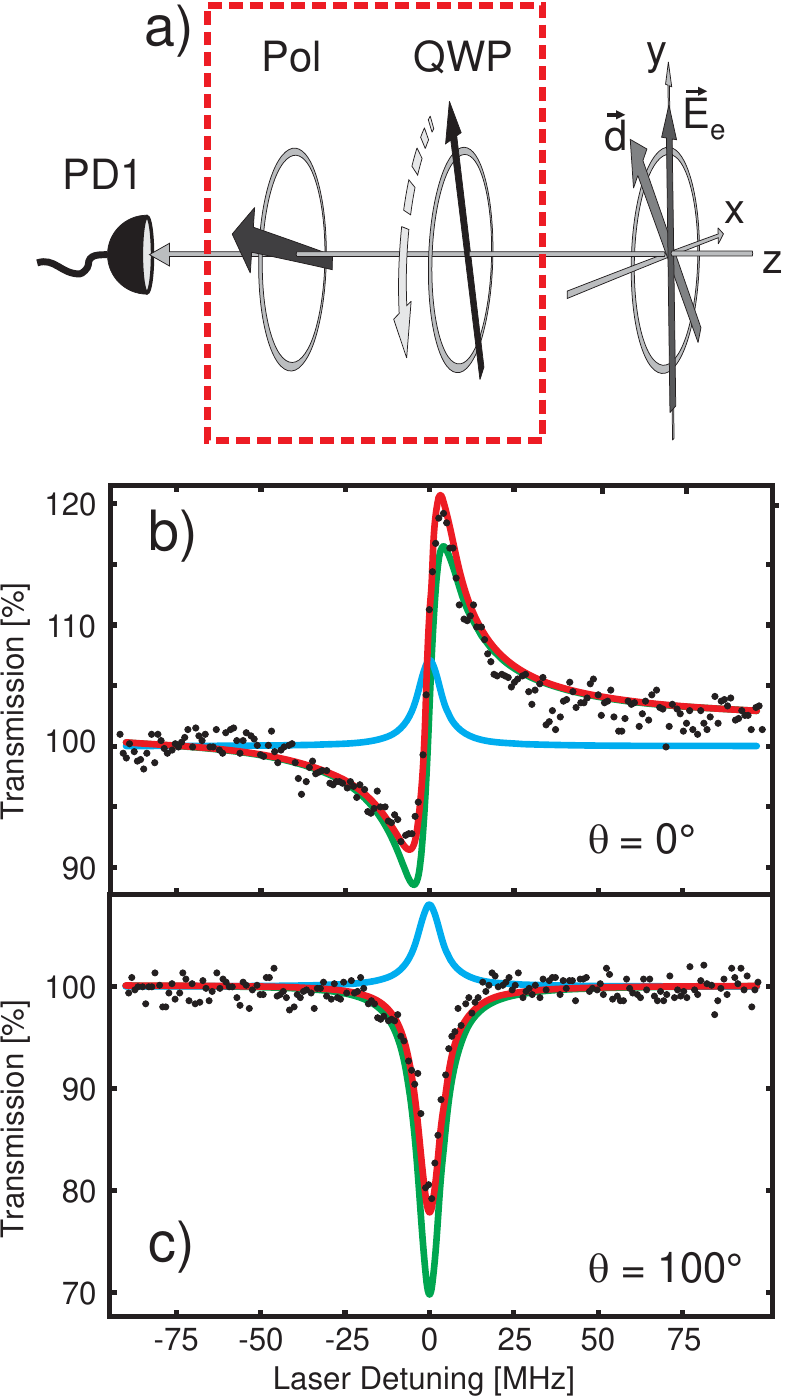}
\caption{ a) Arrangement of a quarter waveplate (QWP) and a
polarizer (Pol) for the separation of the coherent and incoherent
parts of the resonance fluorescence. The laser field
($\mathbf{E}_{\rm e}$) is taken to be along the y-axis whereas the
molecular dipole moment ($\mathbf{d}$) and the QWP are set at $45$
and $\theta$ from this axis, respectively. b, c) Two examples of the
recorded spectra are shown by the symbols. The red curves are the
fits to the experimental data while the blue and the green lines
plot the contributions of the molecular fluorescence intensity and
extinction components respectively, as extracted from the series of
measurements described in the text. \label{polarization}}
\end{figure}

As illustrated by Fig.~\ref{polarization}a, for this experiment we
have introduced a quarter waveplate and a polarizer in the detection
path of PD1 at the position marked by the dotted red box in
Fig.~\ref{setup}c. We set $\mathbf{E}_{\rm e}$ at $45^\circ$ with
respect to the transition dipole moment of the molecule and at
$80^\circ$ with respect to the axis of the polarizer. After
traversing the quarter waveplate, the amplitudes and phases of
$\mathbf{E}_{\rm e}$ and $\mathbf{E}_{\rm m}$ have evolved
differently, thus changing the relative phase $\psi$ and the
coefficients $A$ and $B$. When rotating the quarter waveplate, both
the strength and the shape of the $B$-Term in Eq.~(2) are altered,
but the Lorentzian peak of the $A$-term only changes its magnitude.
By recording and analyzing a series of spectra, we were able to
identify each component in a unique manner. The symbols in
Figs.~\ref{polarization}b and c display two examples of such an
analysis. The red curve is the fit to the experimentally recorded
spectrum. The blue and the green spectra show the molecular
fluorescence intensity and the extinction components that were
extracted from the total signal, respectively. In case of
Fig.~\ref{polarization}b, the addition of the $A$-term has resulted
in a final asymmetric dispersive shape whereas in
Fig.~\ref{polarization}c, the peak of $8\%$ associated with the
$A$-term has reduced a $30\%$ dip of the $B$-term to yield a total
observable dip of $22\%$.

From the knowledge of the extinction $B$-term and $I_{\rm d}$ we can
calculate the absolute number of the coherently scattered photons
for a given excitation intensity. To measure the intensity
dependence of the coherent resonance fluorescence, we have varied
the excitation intensity between 5~pW and 10~nW. The black symbols
in Fig.~\ref{coh-incoh} show the power of the coherent emission of a
single molecule, extracted according to the procedure described
above. We obtain an excellent agreement between the measured and the
theoretical data (black solid curve) without using any free
parameter. The red dots in Fig.~\ref{coh-incoh} show the
conventional saturation curve obtained from the fluorescence
excitation signal recorded simultaneously on PD2 as a direct measure
of the excited state population, also in excellent agreement with
the function $S/(1+S)$ displayed by the solid red curve. We note
that in our work $S$ was determined independently by analyzing the
measured linewidths of fluorescence excitation spectra, whereby
$S=1$ corresponds to a power broadened FWHM linewidth of
$\sqrt{2}\gamma_0$.

The coherent part of resonance fluorescence has been previously put
to evidence using a heterodyne measurement on the emission of a
single ion~\cite{Hoffges:97}. However, to the best of our knowledge,
a study of its intensity dependence presented in
Fig.~\ref{coh-incoh} has not been reported before. These
measurements are particularly challenging in the solid state because
of the omnipresence of background scattering at the excitation
wavelength. The key to success in our experiment has been the
efficient coupling of the laser light to the molecule, allowing us
to optimize the ratio between the molecular signal and the
background scattering. This is also nicely reflected in the fact
that the onset of saturation at $S=1$ could be achieved at the very
low incident laser power of 350~pW.

\section{Measurement of the Mollow triplet}

The data in Fig.~\ref{coh-incoh} clearly show that the coherent
resonance fluorescence diminishes under strong excitation. This
means that the interference between the laser beam and the molecular
emission is reduced, and the dip in the transmission spectra
recorded on PD1 gradually disappears. In parallel to this process,
the inelastic component of the resonance fluorescence is expected to
increase and give rise to a triplet spectrum as first described by
B. Mollow in 1969~\cite{Mollow:69}. The so-called Mollow triplet has
been indeed examined in various gaseous
systems~\cite{Wu:75,Walther:05}, but its direct observation has
turned out to be elusive in the solid state.

\begin{figure}[b!]
\centering
\includegraphics[width=8 cm]{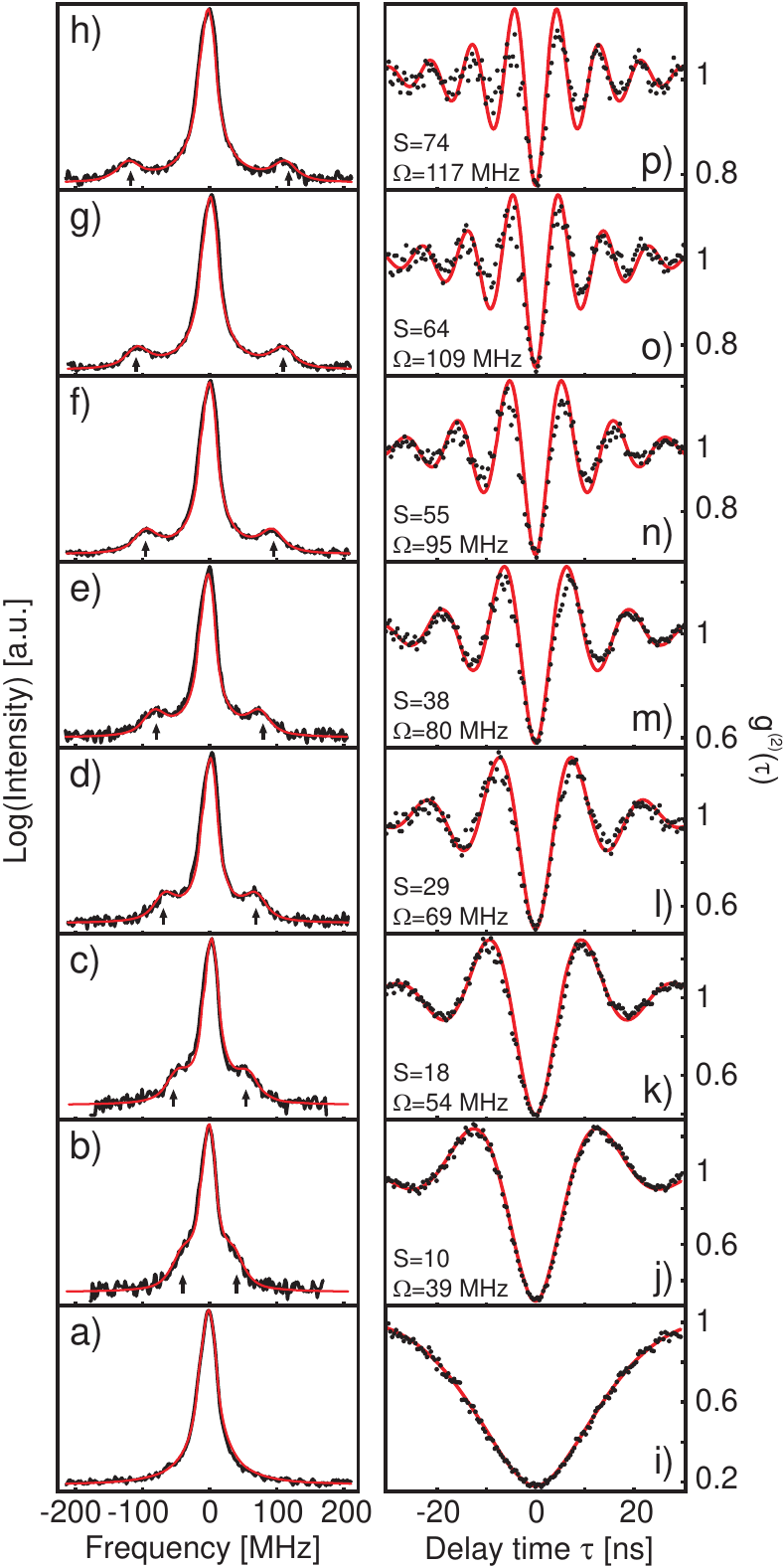}
\caption{a) Transmission of the FPC under laser illumination. b-h)
Series of resonance fluorescence emission spectra (black) recorded
by scanning the FPC for increasing excitation intensities. The red
curves plot the calculated spectra. i) Second-order autocorrelation
function of the single molecule under weak excitation, displaying
photon antibunching. j-p) Series of autocorrelation functions
recorded at the same time as the spectra shown in b-h). The Rabi
frequency $\Omega$ extracted from the fit (red curves) to the
$g^{(2)}(\tau)$ data is indicated in each figure together with the
corresponding saturation parameter $S$. \label{Mollow}}
\end{figure}

To record and study the incoherent part of the spectrum, we used the
detection path in reflection that leads to PD3 in Fig.~\ref{setup}c.
The fluorescence was sent through a short-pass filter to cut out the
Stokes shifted fluorescence corresponding to the $2\rightarrow3$
transitions, while transmitting light at $\lambda_{21}$ and part of
the broad phonon wing of the zero phonon line. Here the angle
between $\mathbf{E}_{\rm e}$ and that of the molecular dipole was
set to $45^\circ$ and the analyzing polarizer was aligned at
$90^\circ$, filtering out the background scattering of the laser
light with a ratio of 300:1. The remaining fluorescence was coupled
into a multimode optical fiber with a core diameter of $50~\rm{\mu
m}$ for convenience of handling. The output of the fiber was sent to
a home-built Fabry-Perot cavity (FPC) with a free spectral range of
356~MHz and transmission of approximately $15\%$. As shown by the
lowest spectrum in Fig.~\ref{Mollow}a, an instrumental linewidth of
about 14~MHz was measured when the reflected laser light was coupled
to the resonator. Despite polarization filtering of the
fluorescence, the background laser scattering was of the same order
of magnitude as the molecular fluorescence and was exploited for
calibrating the frequency axis of each FPC scan.

Figures~\ref{Mollow}b-h show a series of emission spectra recorded
by exciting a molecule on resonance, collecting its fluorescence
close to the same wavelength in the fashion described above and then
scanning the FPC several times, equivalent to an integration time of
3~s per scan pixel. As the excitation intensity was raised (moving
upward in the figure), side bands appeared in the molecular emission
spectrum with increasing frequency separations. The appearance of
the Mollow triplet can be understood as the result of the periodic
modulation of the density matrix elements and thus the molecular
polarization at the Rabi frequency.

To obtain an independent measure of the Rabi frequency, we split the
Stokes-shifted fluorescence beam on PD2 via a beam splitter onto two
photodetectors as sketched in the blue dotted box in
Fig.~\ref{setup}c. By using the common start-stop
scheme~\cite{HanburyBrown:56}, we recorded the intensity correlation
function $g^{(2)}(\tau)$~\cite{Loudon} for each measurement
presented in Figs.~\ref{Mollow}b-h. The results are plotted in
Figs.~\ref{Mollow}j-p. Figure~\ref{Mollow}i confirms that as
expected, we observed pure antibunching at low intensities. However,
as the excitation was made stronger, $g^{(2)}(\tau)$ underwent
oscillations~\cite{Basche:92b}. By fitting the curves in
Figs.~\ref{Mollow}i-p, we determined the Rabi frequencies $\Omega$
in each case. These values together with the corresponding
saturation parameters $S$ are indicated in Figs.~\ref{Mollow}j-p.

Using the independently measured values of $\Omega$, the laser
background and the FPC line profile (see Fig.~\ref{Mollow}a), we
then calculated the theoretical spectra for the Mollow triplet shown
by the red curves in Figs.~\ref{Mollow}b-h. The agreement between
theory and experiment is excellent with only one scaling parameter
for the signal intensity. To our knowledge, this constitutes the
first observation of the Mollow triplet from a single emitter in the
solid state, further confirming the potential of these systems for
well-understood and controlled quantum optical experiments.

\section{Single molecule detection with ultra-faint light sources}

\begin{figure}[b!]
\centering
\includegraphics[width=8 cm]{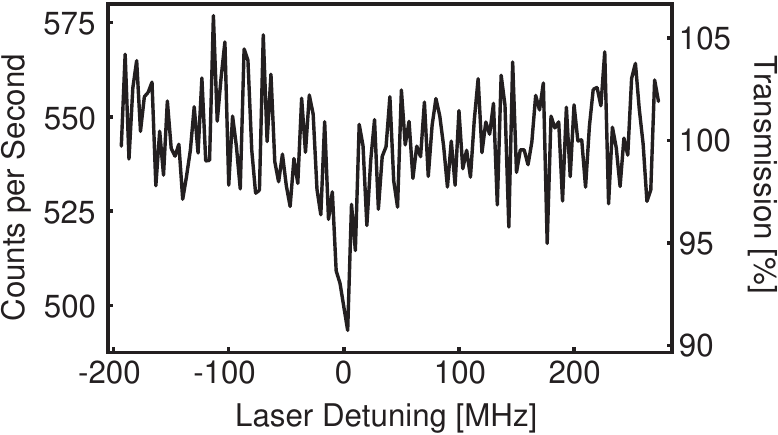}
\caption{An extinction spectrum recorded from a single molecule
under an ultra-faint excitation power of 550 photons per second.
\label{few-photon}}
\end{figure}

Commonly a high flux of photons is required in spectroscopy to
achieve a single excitation because either the observed transitions
are strongly broadened or the light beam is focused very weakly.
Moreover, a certain minimum excitation rate is required to obtain
fluorescence count rates larger than the square root of the detector
noise level. A remarkable advantage of coherent detection of single
molecules, however, is that the detected intensity can easily be
larger than the detector noise even if the molecular emission is
much weaker. This is illustrated in Fig.~\ref{few-photon} where a
single molecule is detected under an incident power of merely 550
photons per second, equivalent to 150 attoWatt. The transmission
spectrum shows the raw data integrated over 4 seconds at each pixel.
When analyzing this result in the context of an interference
phenomenon, we find that the dip of 50 counts per second originates
from a detected molecular emission of only 1.1 photons per second,
which is about 10 times lower than the noise associated with the 150
dark counts of our detector. Thus, the high signal contrast of
coherent extinction measurements is ideally suited for detecting
weak emitters such as single rare earth ions, especially if one
would exploit additional improvements in the SNR via lock-in
detection.

The ability to detect a single emitter with an ultra-faint
illumination also shows the feasibility of exciting a single
molecule by the emission of a second molecule acting as a source of
Fourier-limited single photons~\cite{Kiraz:04}. Considering that the
fluorescence of a single molecule typically amounts to about $10^5$
detected photons per second, we can easily expect a few thousands of
photons per second at the position of the second molecule after
various diagnostics and handling optics. Indeed, calculations show
that in the case where an emitter is illuminated by a single photon
Gaussian wavepacket of width $\gamma_0$, there is a higher than
$50\%$ chance for it to act as a $\pi$-pulse to coherently transfer
the population of the ground state to the excited
state~\cite{Domokos:02} or to experience a substantial phase shift
via its interaction with the molecule. Furthermore, it has been
suggested that the molecule can act as a nonlinear medium to couple
two ultra-faint photon fields impinging on it at the same
time~\cite{Domokos:02}.

\section{Conclusion}

By confining a laser beam close to the diffraction limit at T=1.4~K,
we have examined for the first time, the \emph{coherent} interaction
of a single organic molecule with light over 9 orders of magnitude
of excitation power. In the weak excitation regime, we have achieved
a large extinction dip of $11.5\%$ and have demonstrated the
detection of a single molecule with an ultra-faint laser source of
150 attoWatt. In the strong excitation regime, our measurements have
allowed us to distinguish the coherent and incoherent parts of
resonance fluorescence and to study their intensity dependencies.
Furthermore, we could directly record the Mollow triplet as a
landmark of the nonlinear interactions between an emitter and laser
light.

From a fundamental point of view, an intriguing question that arises
is whether one could achieve a perfect coupling between a
\emph{freely} propagating optical mode and a dipolar emitter in a
single-pass encounter, i.e without high-finesse cavities. When
considering the interference nature of the extinction signal, it is
clear that the annihilation of a photon in the forward direction
requires perfect overlap between its mode and that of a dipolar
radiator. Clearly, this condition is not satisfied for a strongly
focused $\rm{TEM}_{00}$ laser beam~\cite{vanEnk:01}. However, it has
been proposed that the combination of an emitter with an axially
oriented transition dipole moment~\cite{pfab:04} and light with a
radially polarized doughnut mode would be
favorable~\cite{Dorn:03,vanEnk:04}. Another exciting possibility
might be to use nano-antennae for engineering the radiation pattern
of the molecule and matching it to the excitation
mode~\cite{Alu:07}. Given the achievements presented in this article
and the simplicity of the SIL technology, we anticipate that our
approach opens doors to the realization of these ideas and a wide
range of other investigations on the coherent and nonlinear
interaction of light with single molecules and ions in the condensed
phase.

We are grateful to A. Renn for experimental support. This work was
financed by the Schweizerische Nationalfond and the ETH Zurich
initiative for Quantum Systems for Information Technology (QSIT).
\\The authors declare that they have no competing financial interests.
\\
Correspondence and request for materials should be addressed to V.S.
(email: vahid.sandoghdar@ethz.ch).

\end{document}